# Chirurgie mini-invasive assistée par ordinateur des tumeurs du foie: spécifications et résultats préliminaires

# Computer-aided hepatic tumour ablation: requirements and preliminary results

## Chirurgie du foie mini-invasive assistée par ordinateur

*David Voirin[1,2], Yohan Payan[1], Miriam Amavizca[1], Christian Létoublon[2], Jocelyne Troccaz[1]*

[1]Laboratoire TIMC - Faculté de Médecine - Domaine de la Merci - 38706 La Tronche cedex - France
[2]Service de Chirurgie générale et digestive - CHU de Grenoble  - BP 217 - 38043 Grenoble cedex 9 - France

Auteur correspondant: Jocelyne Troccaz, Laboratoire TIMC - Faculté de Médecine - Domaine de la Merci - 38706 La Tronche cedex - France, tel: +33 (0)476 54 95 08, fax: +33 (0)476 54 95 55, jocelyne.troccaz@imag.fr

Résumé: La résection de tumeurs hépatiques n'est pas toujours possible; en particulier l'état du patient ou la localisation de la tumeur peuvent être des facteurs empêchant l'abord chirurgical. Des techniques alternatives ont été développées; elles consistent à utiliser localement des agents physiques ou chimiques permettant de détruire la région tumorale. L'hyperthermie par radiofréquence ou la cryothérapie permettent ainsi un abord percutané de la tumeur. Cependant, ces techniques mini-invasives nécessitent une localisation précise et un suivi de la position de la tumeur pour la réalisation du geste. L'utilisation de systèmes de chirurgie assistée par ordinateur peut améliorer la précision d'un tel geste tout en préservant son caractère mini-invasif. Ce papier présente les principes d'un système de destruction mini-invasive de tumeurs du foie assistée par ordinateur. Il décrit de premières expérimentations réalisées, ayant trait à l'évaluation de techniques de mise en correspondance de données. Afin d'être au plus près du protocole chirurgical, les modalités utilisées sont une imagerie pré-opératoire (scanner X ou IRM) et l'imagerie échographique per-opératoire.

Mots clés:
Chirurgie du foie, destruction de tumeurs, chirurgie assistée par ordinateur, recalage d'images, IRM, échographie.

Abstract: Surgical resection of hepatic tumours is not always possible depending on different factors among which their location inside the liver functional segments. Alternative techniques consist in locally using chemical or physical agents to destroy the tumour. Radio-frequency and cryosurgical ablations are examples of such alternative techniques which may be performed percutaneously. This requires a precise localisation of the tumour placement during ablation. Computer-assisted surgery tools may be used in conjunction to these new ablation techniques to improve the therapeutic efficiency whilst benefiting from minimal invasiveness. This paper introduces the principles of a system for computer-assisted hepatic tumour ablation and describes preliminary experiments focusing on data registration evaluation. To keep close to conventional protocols, we consider registration of pre-operative CT or MRI data to intra-operative echographic data.

Version abrégée

1. Introduction

Les Gestes Médico-Chirurgicaux Assistés par Ordinateur (GMCAO) ont connu un fort développement durant la dernière décennie. Visant d'abord à améliorer la précision du geste chirurgical sur la base de données multi-modales afin de le rendre plus efficace et moins iatrogène, les GMCAO obéissent au concept suivant: guider le geste chirurgical en temps réel en contrôlant en per-opératoire que le déroulement est conforme à la stratégie pré-opératoire planifiée. Ces techniques ont initialement été développées pour des applications concernant des tissus non déformables et immobiles (chez un patient anesthésié). L'orthopédie est un domaine privilégié. Les GMCAO se sont plus récemment intéressés aux tissus mous (cœur et foie notamment). Ces derniers présentent une double difficulté.  Mous, non seulement ils se déforment mais encore ils bougent. En ce qui nous concerne, nos travaux ont porté sur le foie. Dans un premier travail, Herline et al. concluent à la faisabilité d'un tel système où les outils chirurgicaux, munis de diodes infrarouges, sont repérés dans l'espace grâce à un localisateur optique. Dans une seconde étude, les mêmes auteurs étudient un processus de recalage basé sur la modélisation de la surface hépatique. L'expérimentation est faite sur fantôme statique. Dans ce papier, nous nous proposons d'aborder ces questions de recalage sur des données de volontaire sain.

2. Matériels et méthodes:

Les GMCAO obéissent à une méthodologie en trois étapes:
- *La planification pré-opératoire:* à partir de données d'imagerie pré-opératoire (tomodensitométrie, IRM), une stratégie pré-opératoire optimale est établie.
- *Le recalage per-opératoire:* cette étape consiste à transposer la planification pré-opératoire durant l'intervention. Dans notre application, des données échographiques repérées dans l'espace grâce à un système 3D permettent ce recalage.
- *Le guidage du geste thérapeutique:* passif, actif ou semi-actif, le guidage intègre en temps réel les données multi-modales du recalage et de la localisation dans l'espace des outils du chirurgien afin de permettre la réalisation effective de la stratégie pré-opératoire optimale.

L'application qui nous concerne obéit à ces trois étapes. Cependant, recalage et guidage se doivent de prendre en compte les deux spécificités du foie: mobilité et déformabilité. Situé au contact du diaphragme, muscle respiratoire principal, le foie bouge sous l'impulsion des mouvements diaphragmatiques. La déformabilité hépatique quant à elle présente deux aspects. Lors du recalage, le foie, tissu mou, peut présenter une forme différente par rapport aux données du planning. Lors du guidage, le geste chirurgical lui–même exerce une pression à la surface du foie susceptible d'entraîner une déformation du foie et un déplacement des structures.

## 3. Expérimentations

### 3.1 Remarques préliminaires

L' objectif principal de cette étude a été d'évaluer la faisabilité d'une assistance informatisée s'appuyant sur les données d'imagerie classiquement utilisées (scanner ou IRM et échographie). Relativement aux autres travaux du même type sur le foie, il s'agissait de tester les possibilités de recalage sur des données aussi proches que possible de la réalité clinique: utilisation de données échographiques acquises mini-invasivement - par opposition à des données de surface acquises par palpation -, test sur des données réelles acquises sur volontaire sain (par opposition à des tests sur fantômes, le plus souvent statiques).

### 3.2 Acquisition des données

Nous avons utilisé un système dont les différents éléments se composent ainsi :
- un échographe portatif (Hitachi EUB 450) dont la sonde est munie de diodes électroluminescentes.
- un localisateur tridimensionnel muni de trois caméras infrarouges (système Optotrak, NDI) permettant de repérer la sonde d'échographie dans l'espace. Ainsi, grâce à l'enregistrement de la position de la sonde simultanément à celui de l'image, les données extraites des images peuvent être représentées dans un référentiel commun donnant ainsi accès à un modèle tri-dimensionnel issu de l'échographie.
- un logiciel informatique permettant de segmenter les images acquises lors d'un examen pré-opératoire (IRM dans notre cas) et lors de l'échographie dite per-opératoire. Ce logiciel permet après segmentation de reconstruire une représentation tridimensionnelle des données segmentées à partir de chacun de ces deux examens (IRM et échographie). Un logiciel de mise en correspondance de données permet enfin de fusionner les deux volumes reconstruits.

Pour l'étape de planification pré-opératoire, nous avons réalisé une IRM chez un sujet sain, cet examen ayant fait la preuve de sa spécificité et de sa sensibilité dans l'exploration hépatique. L'acquisition a été réalisée en mode T2 UTSE selon deux plans orthogonaux : horizontal et frontal. Un trigger respiratoire constitué d'une ceinture mesurant les mouvements d'ampliation thoracique déclenchait l'acquisition en période de fin d'expiration. La distance inter-coupes était de 0,4 cm, l'épaisseur de coupe de 4 mm. Le nombre total de coupes s'élevait à 67. Pour l'étape du recalage per-opératoire, nous avons utilisé l'échographie. La représentation tridimensionnelle échographique obtenue grâce à la localisation de la sonde par localisateur a nécessité plusieurs mises en apnée successives du volontaire. Les clichés ont été enregistrés au cours de la même phase du cycle respiratoire que celle durant laquelle avait lieu l'acquisition IRM en mode trigger. Les coupes réalisées ont été sagittales, horizontales et obliques. Au nombre de 36, elles ont balayé la totalité du volume du foie. Les coupes IRM et échographiques ont été segmentées manuellement.

### 3.3 Tests de recalage

Les représentations tridimensionnelles reconstruites à partir de chacun des examens (échographie et IRM) ont ensuite été fusionnées selon un recalage rigide. Il s'agit de déterminer les 6 paramètres de transformation (rotation et translation) permettant de superposer les deux ensembles de données en minimisant l'erreur aux moindres carrés. La précision du recalage a été estimée en utilisant plusieurs types de tests: tests de stabilité et tests de cohérence. La difficulté principale de cette évaluation résidait dans l'absence d'"étalon or" servant de mesure de référence. Travaillant sur le sujet sain vivant, nous avons été obligés de procéder à une évaluation indirecte de la précision par analyse de la cohérence de différents recalages obtenus sur des données distinctes.

Nous avons préalablement scindé les données IRM pour obtenir deux représentations tridimensionnelles, à partir des coupes axiales pour l'une, des coupes frontales pour l'autre. Différents recalages ont ensuite été réalisés, à plusieurs reprises pour chacun d'eux, entre échographie et IRM frontale, échographie et IRM axiale et IRM axiale et IRM frontale. Afin d'évaluer la stabilité des résultats, chaque matrice de transformation de chaque recalage a été comparée pour chaque essai avec les autres matrices de transformation du même recalage; les recalages ont enfin été combinés par produit matriciel pour vérifier la cohérence des résultats de ces différents recalages.

3.4 Tests de déformation locale

Une seconde série d'images échographiques a également été acquise afin d'évaluer localement la compressibilité du foie. Pour une position donnée de la sonde échographique connue grâce au localisateur, plusieurs images étaient enregistrées pour des valeurs croissantes de pression de la sonde sur le corps du volontaire. Des structures anatomiques de référence ont été segmentées sur ces images et superposées dans un référentiel commun. Leur déplacement résultant de la compression par la sonde échographique a été mesurée.

3. Résultats et discussion

3.1 Recalage

Pour les différents recalages, la distance moyenne résiduelle entre les deux représentations tridimensionnelles à superposer était après recalage de l'ordre de 3 mm. La stabilité des transformations s'avère très satisfaisante. Pour ce qui concerne le test de cohérence, si T1, T2 et T3 sont les matrices de transformation respectivement obtenues pour les recalages "IRM frontale - IRM axiale", "échographie - IRM frontale" et "échographie - IRM axiale", nous devrions, en théorie, obtenir T1*T2=T3. En pratique, nous obtenons une erreur cumulative d'une amplitude d'environ 6 mm mesurée à l'origine du repère échographique situé presque au centre du volume hépatique. Ceci semble tout à fait convenable eu égard aux différentes sources d'erreur. Les principales sources d'erreur nous paraissent être les suivantes:

- précision du trigger respiratoire lors de l'acquisition IRM
- précision de la mise en apnée lors de l'acquisition échographique : le sujet volontaire a dû se mettre plusieurs fois en apnée en essayant de reproduire l'apnée au même temps du cycle respiratoire que lors de l'acquisition IRM
- bouger du volontaire lors des acquisitions
- précision des calibrages de l'IRM (échelle) et de l'échographie (calibrages intrinsèque et extrinsèque)
- précision de la segmentation des données.

3.2 Déformabilité locale

L'approche qualitative basée sur l'observation visuelle des coupes échographiques et des représentations tridimensionnelles a permis de constater qu'une pression abdominale pouvait déformer la face antérieure du foie (face superficielle) sans modifier la face postérieure (face profonde). Il a par ailleurs aussi été observé qu'une déformation du foie dans le sens antéro-postérieur pouvait s'accompagner d'un mouvement de translation. Sur le plan quantitatif, les déformations internes au foie mesurées sur les images échographiques pour une position donnée de la sonde et une acquisition avec ou sans pression ont été de 1,5 cm en coupe horizontale et de 0,7 cm en coupe sagittale. Il est à noter cependant que les pressions exercées étaient très largement supérieures à ce qui est strictement nécessaire à l'acquisition des images. Nous pensons que ces déformations seront négligeables dans des conditions normales d'acquisition échographique.

3.3 Applicabilité de l'approche

Nous avons étudié la faisabilité d'une assistance informatisée pour la destruction mini-invasive de tumeurs hépatiques. Les premiers résultats obtenus sont très encourageants mais la précision ainsi que l'applicabilité clinique se doivent d'être améliorées. Cette étude est donc poursuivie dans les directions suivantes.

- L'utilisation de la surface du foie comme structure de recalage rend l'acquisition de données per-opératoires longue, voire fractionnée en plusieurs sous-acquisitions. Ceci, outre l'inconvénient intrinsèque relatif à un allongement de la procédure opératoire, résulte en une source d'erreurs. Nous souhaitons donc étudier l'utilisation d'autres structures pour le recalage, notamment l'arborescence portale.
- Afin d'améliorer la précision du système, un recalage élastique des données sera utilisé dans les versions à venir.
- Enfin, la modélisation biomécanique par éléments finis du foie et de son environnement anatomique, qui est en cours, permettra une meilleure prédiction des déformations et déplacements de la tumeur. Ceci sera utilisé à la fois pour le choix de la trajectoire d'abord de la tumeur ainsi que pour la mise en œuvre du guidage du geste robotisé ou non.

# Texte principal

## 1. Introduction

Surgical resection of hepatic tumours is not always possible depending not only on their location inside the liver functional segments but also on the patient general state and pathology. Alternative ablative techniques consist in locally using chemical or physical agents to destroy the tumour. These techniques include ethanol injection, cryosurgery, microwave coagulation, interstitial laser photocoagulation and radiofrequency ablation. Each of these techniques have their own advantages, drawbacks and indications (see for instance [1]). The intra-operative monitoring of the ablation is generally performed using intra-operative ultrasounds, radiology or interventional CT or MRI. These procedures may be executed using a laparoscopic access or percutaneously thus reducing the intervention invasiveness. Nevertheless, some clinical publications [2] highlight that percutaneous access procedures seem to be less efficient probably because of a more difficult localisation of the tumour. This paper proposes to develop computer-assistance for improvement of local ablative techniques. The general idea is to separate intra-operative imaging from the monitoring of tool position. For this, intra-operative imaging is used, in a first stage, for registration to pre-operative data. Then, in a second stage, the localization of the surgical tool is computer controlled with no further image acquisition.

Over the last fifteen years, most of the computer-assisted surgery systems were developed in the context of specialities dealing with bony structures or with structures behaving with limited deformations and movements. For example, the so called "navigation systems" which aim is to help the surgeon to know the exact position of the instruments relatively to anatomical structures, were used for orthopaedics (see for instance [3]), neurosurgery, ENT surgery or plastic surgery. Very little attention has been given to computer-assisted soft tissues surgery and even less to computer-assisted liver surgery. Liver surgery simulation [4] and classical laparoscopic[1] techniques are out of the scope of this paper. To our knowledge, research work in the field of computer-assisted liver surgery has been published only very recently. The main difficulty is that the liver position depends both on the patient position and on the point in the respiration cycle. Moreover, the liver moves and deforms during surgery. [5] mostly focuses on the evaluation of liver motion and deformation during respiration. In this perspective, markers on a porcine liver and anatomical landmarks on a patient liver were tracked and measurements were performed respectively during insufflation and respiration. Measurements on patients show a sinusoidal motion with an average amplitude of 10.8mm during respiration and the authors conclude that computer-assisted liver surgery should be feasible; these numbers are constraints to be used in the system design. [6] focuses on the evaluation of surface registration techniques. The idea was to register pre-operative CT data to surface points intra-operatively collected on the liver surface whilst overcoming the inaccuracy of point-to-point registration associated to landmarks matching. The evaluation was performed on a static phantom liver.

## 2. Materials and methods

### 2.1 The approach

The general approach consists in the following steps:

---

[1] Even if the laparoscope is held and moved by a robot.

- A pre-operative model of the liver is reconstructed from CT or MRI data, integrating the main reference anatomical hepatic structures as well as the target tumours.
- The surgeon uses this model to define a planning to reach the tumour (entry points, spatial targets and safe trajectory with anatomical structures that should be avoided)
- Just before surgery, intra-operative ultrasonic images are collected to get a set of 3D points located onto the references hepatic structures. Note here that the ultrasonic modality has been preferred to other intra-operative ones because it is a very common, safe and relatively inexpensive imager as compared to radiology or interventional CT or MR.
- This set of 3D points is matched onto the pre-operative model of the liver, by the mean of a registration technique. The matching transformation is then used to transfer the pre-operative planning into the operative room.
- Finally, the surgical tool position is computer-controlled to guarantee the execution of the planned strategy. No further image acquisition is needed for this guiding phase.

The three main stages, namely planning, registration and guidance, are described below in more details.

2.1.1. Planning phase

Pre-operative 3D data are collected. In these 3D data, two types of relevant anatomical structures have to be segmented. Structures that will participate to the planning - tumour and structures at risk such as vessels or bile ducts - are called "planning structures". Taking into account the information provided by these structures, the planning phase allows the selection of a needle trajectory and a target position. Structures which will be used for registration are called "reference structures": e.g. the liver surface. 3D representations of these two types of structures are obtained in the pre-operative referential. Let us mention that a structure can belong both to the planning and to the reference structure classes.

2.1.2. Registration phase

Ultrasound data of the reference structures are collected just before the intervention. The echographic probe is equipped with localisation features - for instance infrared diodes - which are tracked in real time using a localizer (see figure 1). Each time that an echographic image is recorded, the 6 position parameters of the probe are also recorded thus localizing the 2D echographic image in the 3D space. We call this device "2.5D echography". Thanks to the image position, the segmented structures are also localized in the 3D space, thus allowing to build a 3D representation of the reference structures in the intra-operative referential. Let us mention that this resulting representation may be a sparse set of data. Indeed, those data are used for registration to mainly compute the geometric transformation between the set of 3D points and the pre-operative model of the liver. Therefore, a complete and homogeneous echographic reconstruction of the whole liver surface is not necessary. At that point, planning data (target position of the tumour and trajectory) can be mapped to the intra-operative conditions using this transform.

2.1.3. Guidance phase

Then, different types of guiding systems [7] may be used to reach the target position through the planned trajectory. A passive system based on surgical instrument tracking capabilities may give information that compare the executed trajectory to the planned one. The surgeon can then use this information to control his action. On the other side, an active robotic system may automatically move the instrument according to the planned trajectory. The choice of one particular type of system will be discussed in section 4.

*2.2. Experiments*

2.2.1. Preliminary remarks

The main purpose of this preliminary work was to evaluate the feasibility of image-based guidance provided that the image modalities are determined by conventional procedures. We mostly focused on registration questions. Our goal was to be able to quantitatively evaluate the algorithms that match *CT with 2.5D echography* in rather realistic conditions. [6] evaluates the surface registration of CT data with palpated points on a static phantom. Registration is tested for several data acquisitions on different parts of the liver surface. Results are compared to a point-to-point registration considered as the gold standard. This work differs from [6] on two points. Firstly, we preferred echographic data acquisition to surface point collection: it allows to envision the use of the system for percutaneous therapeutic actions whilst surface point collection does not. Secondly, we decided to work on real data acquired with a volunteer. In this case, no gold standard is available. Section 2.2.3 is going to show how accuracy could be characterized however.

In the first stage of this study, it was assumed that the key steps of the protocol, namely image acquisitions and guidance, can be executed at a same point in the respiration cycle, for instance during expiration. This assumption will be discussed in section 4.

2.2.2 Data acquisition

Because the experiments were conducted with a volunteer, it was not reasonable to use CT acquisition. Therefore, MRI was selected for pre-operative modelling of the liver. The MRI system is a Philips Gyroscan ACS-NT 1.5Tesla system. The acquisition was synchronized with the respiration cycle and was done only during the expiration phase. Two T2 UTSE sequences were performed resulting in one axial and one frontal imaging data (see figure 2). Inter-slice distance is 4mm; slice thickness is 4mm. The total number of slices is 67. The external surface of the liver was chosen as the "reference structure" and was manually segmented onto both MRI data. Two 3D representations consisting in two sets of 3D points were thus obtained: they are named $MR_{axial}$ and $MR_{frontal}$ and visualized on figure 3.

In a second stage - that simulates intra-operative procedures - 2.5D echographic acquisitions were performed. A first set of data was acquired during the apnoea that follows expiration. On this first set of images, the liver surface was also manually segmented. Figure 4 shows a typical echographic image of the liver. These segmented images provided a 3D representation named "Echography representation" (see also figure 4). The echographic system was a HITACHI EUB450 with a 3.5MHz probe. The infra-red based localizer was an Optotrak system from Northern Digital Inc.

Then, a second set of echographic data was acquired. For different positions of the echographic probe, several images were recorded and localized while applying an increasing pressure on the body with the probe. The maximum exerted pressure was much higher than usually necessary for data acquisition. In order to roughly evaluate liver local deformability, specific structures of interest that correspond to a single position were manually segmented on the different images.

2.2.3 Surface registration testing

The registration algorithm is a surface matching using a distance map recorded in an octree spline data structure (cf. [8]). This data structure is computed from the most dense representation, namely the pre-operative MRI model in our case. The sparse representation should be a subset of the dense one. The algorithm iteratively moves the sparse representation relatively to the dense one and computes the parameters that minimize the distance function between the two representations. The optimisation uses the Levenberg-Marquardt procedure. Thanks to the distance map, computations are very fast. At a starting point, only rigid matching has been used with a computed transformation that corresponds to the 6 position parameters. Intrinsic deformations of the liver, due to its elastic structure, were therefore not compensated by the registration. Because data were acquired in conditions where no gold standard was available - in other words, it was not possible to know the exact transform between MR and echographic data - two types of tests were used to evaluate the registration.

The first test named "repeatability test" consists in running the registration algorithm from several initial relative positions of the MRI and echographic representations and to observe the repeatability of the computed transform. Such a test informs on the presence of local minima in the vicinity of the solution. The value of the residual mean square (rms) after registration is also an indicator of registration accuracy: a large rms would mean an inaccurate registration or mismatched data (this may occur when the sparse data are not a subset of the dense ones). Even if we are mostly interested in registering echography to MRI, we considered for this test three registrations: Echography to $MR_{frontal}$, Echography to $MR_{axial}$ and $MR_{frontal}$ to $MR_{axial}$.

The second test named "closed-loop accuracy test" takes advantage of the two MR acquisitions. Starting from the three registrations mentioned above (Echography to $MR_{frontal}$, Echography to $MR_{axial}$ and $MR_{frontal}$ to $MR_{axial}$), we obtained respectively the transforms $T_{echo\_frontal}$, $T_{echo\_axial}$ and $T_{frontal\_axial}$. Let us notice that a transform $T_{R1\_R2}$ applied to a point or vector given in a referential $R_1$ maps it to $R_2$. If the three registrations were perfectly accurate and if there was no systematic error coming from data acquisition and structure segmentation, we would have (see figure 5):

$$T_{frontal\_axial} * T_{echo\_frontal} = T_{echo\_axial}$$

3. Results

*3.1 Repeatability test*

Table 1 shows the obtained results for five initial positions; a transform is represented as one translation vector and 3 rotation angles. It can be seen that the repeatability is very good. The typical value of the rms we got is about 2mm which is very reasonable considering the different sources of errors listed in section 4. Figure 6 and 7 respectively show the Echography representation superimposed to the $MR_{frontal}$ one and the $MR_{frontal}$ representation superimposed to the $MR_{axial}$ one after registration.

### 3.2 Closed-loop accuracy test

Since the repeatability test gave good results, the closed loop accuracy test was performed on a single set of combined registration data. Table 2 gives the values of $T_{echo\_frontal}$, $T_{echo\_axial}$, $T_{frontal\_axial}$ thus obtained and compares $T_{frontal\_axial} * T_{echo\_frontal}$ to $T_{echo\_axial}$. The homogeneous coordinates notation is used. Applied to the origin of the echographic referential, these two transformations give the following error vector (6, -1.84, -1.53). Let us remind that it is a cumulative error expressed in millimetres.

### 3.3 Local deformability

For each set of pressure-varying echographic images, some relevant structures were segmented and their displacement measured as the pressure of the probe was significantly increased. Since the images were localized in a common referential, it was possible to superimpose segmented data and to evaluate local deformations. Figure 8 illustrates this evaluation for a pseudo-sagittal echographic acquisition. As it can be seen on the figure, the displacement of the posterior part of the liver in contact with the vena cava remains very small when the pressure increases whilst the anterior part of the liver is significantly deformed. Moreover, a translation in the cranio-caudal direction can be detected. Considering that the pressure required for image acquisition is much lower than what was applied for these experiments, we can reasonably expect that the posterior structures of the liver do not move because of the image acquisition but simply because of respiration.

## 4. Discussion

### 4.1 Accuracy issues:

Many sources of errors can be mentioned to explain our results:

- MRI acquisition: On one hand, the respiration triggering was not perfect and some data were acquired during inspiration resulting in errors in the MR representation of the liver. On the other hand, the slice thickness (4mm) introduces some partial volume effect resulting also in inaccuracies.
- Echographic acquisition: on one hand the volunteer had to stop breathing during acquisition. Because the acquisition was long, several sequences were necessary to build the complete representation. The volunteer tried to reproduce apnoea at the same point in respiration but these different sequences result in errors in the echographic representation. On the other hand, the pressure of the probe on the body for image acquisition may result in some tissue local deformation which also could introduce error in the echographic representation of the liver surface.
- MRI and echographic calibration: the image parameters (scale, mm/pixel ratios and geometric relationship between the probe and the localizing features) determined by calibration procedures may introduce errors. The typical rms values after calibration are about 1mm for echographic acquisition. Moreover, each of these imaging modalities may result in distorted representations of the organs. Modelling of these distortions has not been integrated to this study.

- MRI and echographic data segmentation: considering the MR pixel/mm ratio of 1.465 and the size and quality of the images, it looks reasonable to consider that a one to 2 pixels error results from manual image segmentation. This corresponds to a 1.5 to 3 mm error.
- Registration: the registration error is directly related to the quality of data. Tests of the registration algorithms performed on rigid phantoms demonstrated sub-millimetre accuracy. In the experiments presented, distortions of the imaging modalities and the fact that the 3D representation were built from several points of the respiration cycle may degrade the results.

Among all those possible sources of errors, we assume that the main inaccuracy certainly comes from the respiration of the volunteer during data acquisition. Different solutions may be used to improve these results. Concerning pre-operative data, MRI acquisition on volunteers has to be replaced for patients by spiral CT acquisition that is much faster. We can reasonably envision one or two CT acquisitions in apnoea after inspiration and expiration with smaller inter-slice distance and slice thickness. This would result in a more accurate representation of the liver geometry. Concerning intra-operative data acquisition, the process was much too long resulting in successive phases of apnoea. It has to be shortened significantly. This would probably mean that some other reference structures should be considered. This has to be carefully studied from the viewpoints of clinical feasibility and registration robustness. During the surgical action, apnoea can be reproduced or movements of the liver might be restrained significantly using jet ventilation, which is a high frequency, low amplitude ventilation.

The registration accuracy could be improved by making use of elastic registration methods that combine rigid transformations with local deformations based, for example, on Spline functions (see [9] for instance). This elastic registration could reduce the effect of image distortions.

*4.2 Clinical applicability*

Two main approaches can be envisioned for action guidance. The first one was adopted for the present work; it considers that the motion of the liver can be annulated thanks to apnoea conditions during data acquisition and needle guidance. The second approach would be to model the liver motion from CT acquisitions in order to predict the position of the liver at each instant of the respiration cycle and to synchronize the surgical action with this motion. In the first case, a navigational assistance can be used to guide the surgical action, but a robot would probably make the action faster and therefore more accurate. In the second case, a robot has to be used to guarantee the accuracy of the therapeutic action.

For each case, as was underlined in section 4.1, the intra-operative data acquisition should be made as fast as possible, not only to improve the system accuracy but also to reduce the intervention time. On one hand, this implies, as we suggested below, that some reference structures other than the whole liver surface should be selected. For instance, an anatomical structure located as close as possible to the tumour position could be chosen. A further study has to be performed on this very important topic. On the other hand, intra-operative image processing tools must be developed to avoid any involvement of the user in these tasks other than as a supervisor. In particular, approaches such as [10] that was developed for US/CT registration of the pelvic region allows to perform simultaneously and automatically registration and segmentation. This approach will be integrated to a further version of this system to suppress user involvement in intra-operative image processing. Pre-operative image segmentation is less critical; meanwhile, approaches such as proposed by [11], could be introduced.

Finally, because needle introduction may involve tissue deformations, these deformations should be evaluated and modelled as precisely as possible. If significant, they will have to be integrated in the planning. Indeed, a trajectory could be preferred to another one when it avoids critical structures and also if deformations are most unlikely to occur. A finite-element model integrating the liver and its anatomical environment is being studied.

## 5. Conclusion

In this paper, experiments on real data of a human liver were presented. The aim was to evaluate the feasibility of image-guided tumour ablation based on conventional imaging modalities: namely pre-operative CT or MRI data and intra-operative echographic images. Because no gold standard was available, we used a closed-loop accuracy test. Data registration appeared to be repeatable and the obtained accuracy - even limited by data acquisition - allows envisioning the clinical applicability of the method. Further work will be undertaken to automate the intra-operative segmentation task, to improve the registration accuracy and practical feasibility and to implement the planning and guiding functions.

## Acknowledgments

This work has been supported by La Fondation de la Recherche Médicale. We thank Patrick Vassal, Pr Lebas, Pr Coulomb, Dr Sengel and Dr Teil for their assistance in the image acquisition processes.

Figures legends

Figure 1: Echographic set-up

Figure 2: MR data

Figure 3: Pre-operative 3D representation

Figure 4: Intra-operative Echography representation

Figure 5: "Closed-loop" accuracy test

Figure 6: Echography and $MR_{frontal}$ representations

Figure 7: $MR_{frontal}$ representation superimposed to the $MR_{axial}$ representation after registration

Figure 8: Local deformability under compression

Figures

*Figure 1: Echographic set-up:* (left) the echographic probe equipped with infrared diodes - (right) the three cameras for diodes localization and tracking and position measurement.

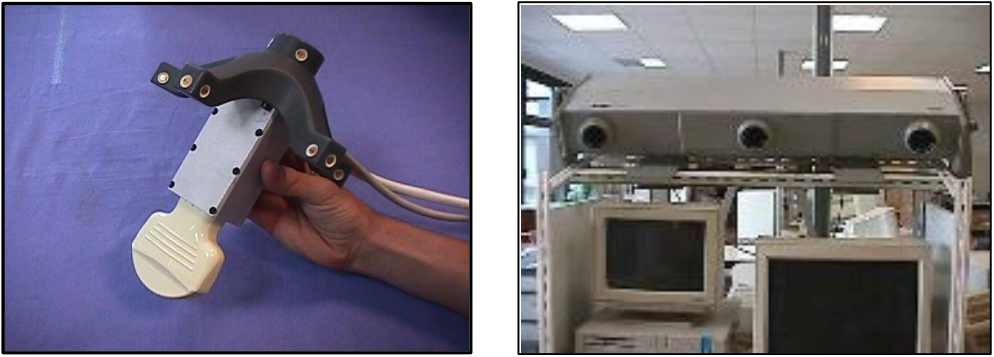

*Figure 2: MR data:* (left) frontal and (right) axial views of the liver

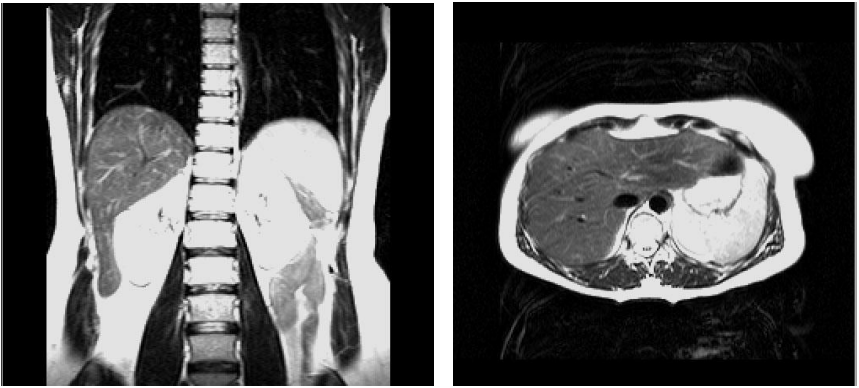

*Figure 3: Pre-operative 3D representation*: (left) $MR_{axial}$ (right) $MR_{frontal}$

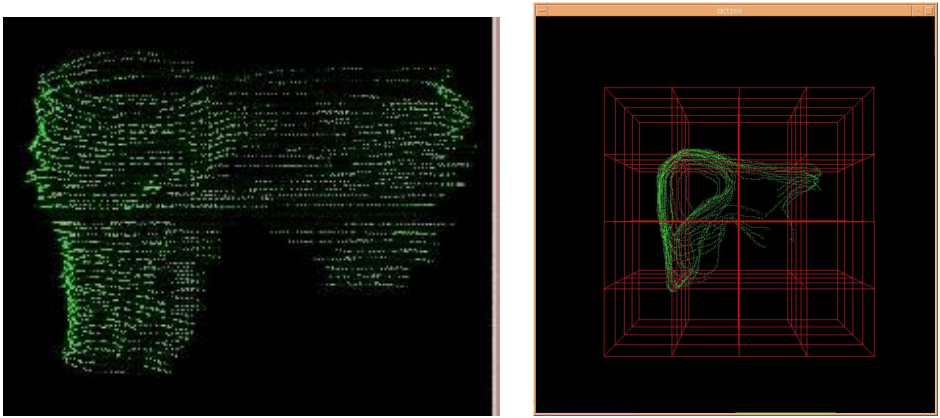

PDF Creator - PDF4Free v2.0                         http://www.pdf4free.com

*Figure 4: Intra-operative image acquisition:* (left) typical liver image - (right) Echography representation

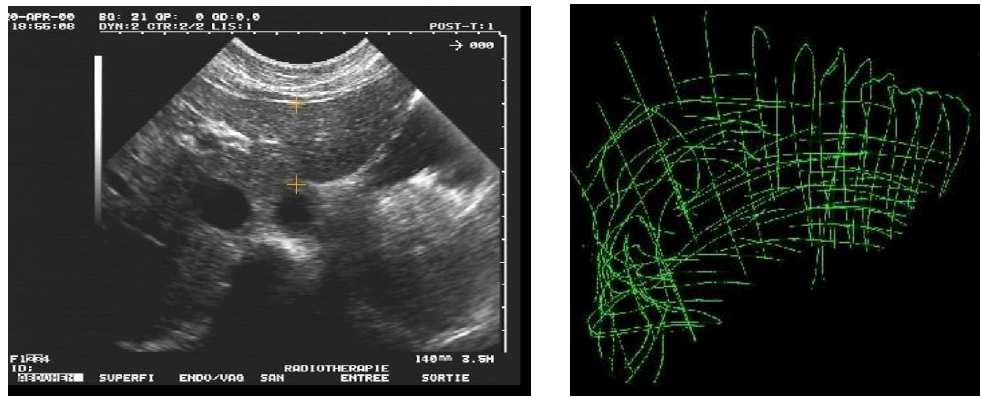

*Figure 5: "Closed-loop" accuracy test*

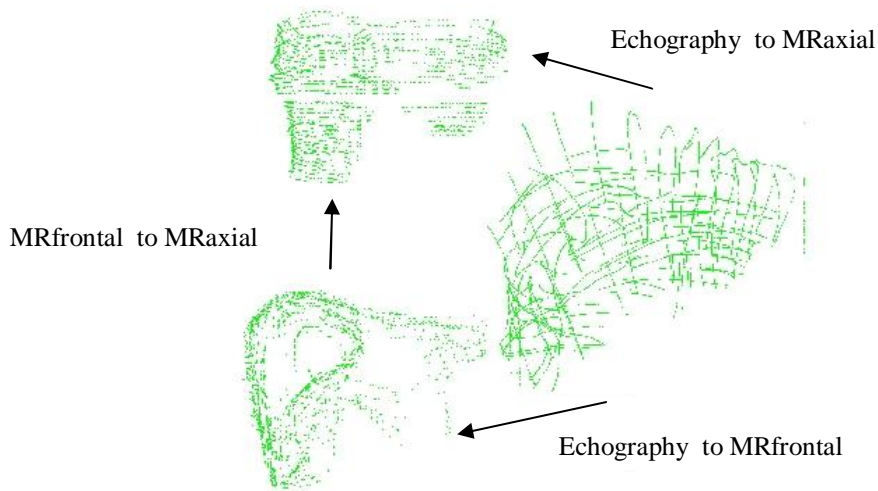

Echography to MRaxial

MRfrontal to MRaxial

Echography to MRfrontal

*Figure 6: Echography (pink) and $MR_{frontal}$ (green) representations*: (left) before and (right) after registration

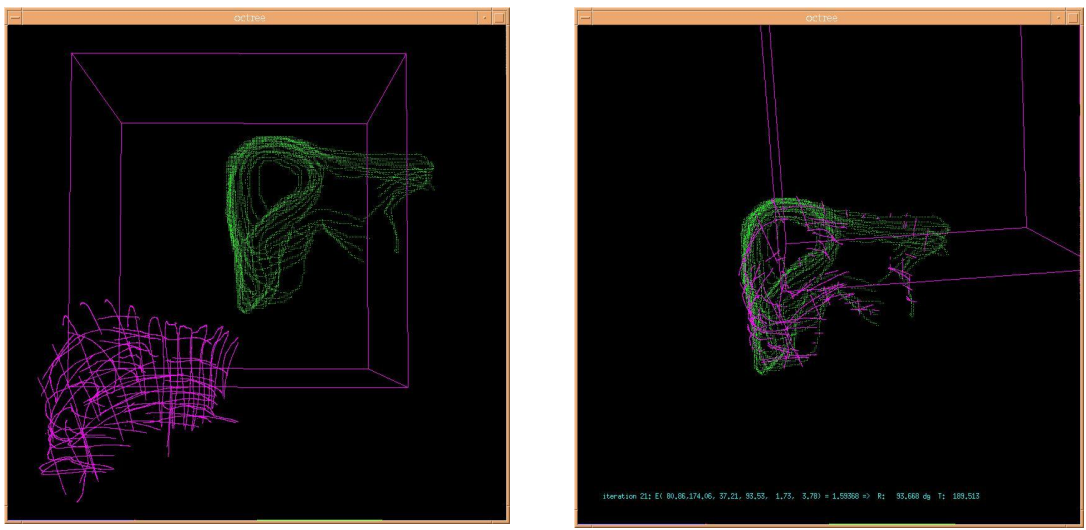

*Figure 7: MR$_{frontal}$ representation (pink) superimposed to the MR$_{axial}$ representation (green) after registration*

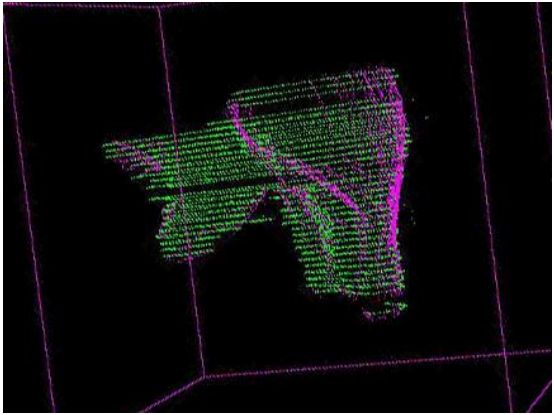

*Figure 8: Local deformability under compression:* (left) low and (middle) high compression - (right) superimposed segmented data.

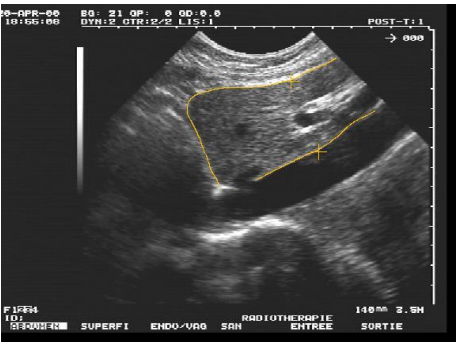 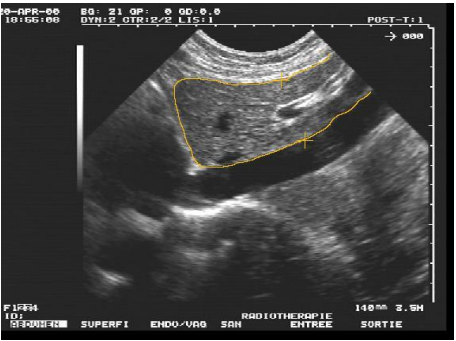 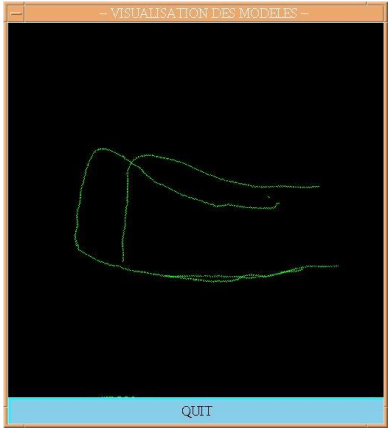

Tables

*Table 1: Repeatability test*

| | Test no1 | Test no2 | Test no3 | Test no4 | Test no5 | mean | sigma | normalized sigma |
|---|---|---|---|---|---|---|---|---|
| **MRaxial to echography** | | | | | | | | |
| Tx | 85,57 | 85,46 | 85,56 | 85,49 | 85,61 | 85,54 | 0,06 | 0,07 |
| Ty | 26,23 | 26,32 | 26,21 | 26,38 | 26,19 | 26,27 | 0,08 | 0,30 |
| Tz | -46,13 | -46,06 | -46,19 | -46,09 | -46,15 | -46,13 | 0,05 | -0,11 |
| Phi | 1,46 | 1,61 | 1,47 | 1,58 | 1,48 | 1,52 | 0,07 | 4,64 |
| Teta | -2,40 | -2,30 | -2,36 | -2,31 | -2,38 | -2,35 | 0,04 | -1,86 |
| Psi | 1,53 | 1,48 | 1,55 | 1,46 | 1,53 | 1,51 | 0,04 | 2,47 |
| **MRfrontal to MRaxial** | | | | | | | | |
| Tx | 3,49 | 3,59 | 3,56 | 3,53 | 3,53 | 3,54 | 0,04 | 1,09 |
| Ty | 125,15 | 125,27 | 125,22 | 125,30 | 125,29 | 125,25 | 0,06 | 0,05 |
| Tz | 9,51 | 9,54 | 9,52 | 9,60 | 9,53 | 9,54 | 0,03 | 0,36 |
| Phi | 84,66 | 84,60 | 84,55 | 84,56 | 84,55 | 84,58 | 0,05 | 0,06 |
| Teta | 5,48 | 5,38 | 5,45 | 5,37 | 5,38 | 5,41 | 0,05 | 0,89 |
| Psi | 5,11 | 5,05 | 5,09 | 5,11 | 5,07 | 5,09 | 0,03 | 0,57 |
| **MRfrontal to echography** | | | | | | | | |
| Tx | 73,07 | 73,25 | 73,28 | 73,21 | 73,27 | 73,22 | 0,09 | 0,12 |
| Ty | 180,25 | 179,92 | 179,95 | 179,82 | 179,94 | 179,98 | 0,16 | 0,09 |
| Tz | 48,37 | 48,21 | 48,25 | 48,39 | 48,28 | 48,30 | 0,08 | 0,16 |
| Phi | 93,84 | 93,99 | 93,92 | 94,09 | 94,04 | 93,98 | 0,10 | 0,10 |
| Teta | 5,89 | 5,78 | 5,77 | 5,73 | 5,75 | 5,78 | 0,06 | 1,11 |
| Psi | 8,16 | 8,06 | 8,09 | 8,13 | 8,09 | 8,11 | 0,04 | 0,46 |

Tx, Ty, Tz : translation parameters
Phi, Teta, Psi : rotation angles

*Table 2: Closed-loop accuracy test*

Tfrontal_axial

| 1,00 | 0,02 | -0,02 | -5,43 |
|---|---|---|---|
| 0,02 | 0,08 | 1,00 | -15,33 |
| 0,02 | -1,00 | 0,08 | 121,43 |
| 0,00 | 0,00 | 0,00 | 1,00 |

Techo_frontal

| 1,00 | -0,02 | 0,03 | 78,21 |
|---|---|---|---|
| 0,03 | -0,07 | -1,00 | 174,20 |
| 0,02 | 1,00 | -0,01 | 29,84 |
| 0,00 | 0,00 | 0,00 | 1,00 |

Techo_axial

| 1,00 | -0,01 | -0,01 | 82,07 |
|---|---|---|---|
| 0,01 | 1,00 | -0,01 | 28,57 |
| 0,01 | 0,01 | 1,00 | -49,28 |
| 0,00 | 0,00 | 0,00 | 1,00 |

Tfrontal_axial*Techo_frontal

| 1,00 | 0,00 | -0,02 | 76,06 |
|---|---|---|---|
| 0,04 | 0,99 | -0,09 | 30,42 |
| 0,02 | 0,15 | 1,00 | -47,74 |
| 0,00 | 0,00 | 0,00 | 1,00 |